\documentclass[10pt,a4paper,prl,showpacs,noshowkeys,twocolumn,superscriptaddress]{revtex4}
\usepackage{times}
\usepackage{amsmath}
\usepackage{amssymb}
\usepackage[dvips]{graphicx}
\usepackage[unicode=true,pdfusetitle,
 bookmarks=true,bookmarksnumbered=false,bookmarksopen=false,
 breaklinks=false,pdfborder={0 0 1}]
 {hyperref}

\begin{document}

\title{Bright solitons from the nonpolynomial Schr\"{o}dinger equation \\
with inhomogeneous defocusing nonlinearities}
\author{W. B. Cardoso}
\affiliation{Instituto de F\'{\i}sica, Universidade Federal de Goi\'{a}s, 74.001-970, Goi%
\^{a}nia, Goi\'{a}s, Brazil}
\author{J. Zeng}
\affiliation{State Key Laboratory of Low Dimensional Quantum Physics, Department of
Physics, Tsinghua University, Beijing 100084, China}
\author{A. T. Avelar}
\affiliation{Instituto de F\'{\i}sica, Universidade Federal de Goi\'{a}s, 74.001-970, Goi%
\^{a}nia, Goi\'{a}s, Brazil}
\author{D. Bazeia}
\affiliation{Instituto de F\'{\i}sica, Universidade de S\~{a}o Paulo, 05314-970, S\~{a}o
Paulo SP, Brazil}
\affiliation{Departamento de F\'{\i}sica, Universidade Federal da Para\'iba, 58.059-900,
Jo\~{a}o-Pessoa, Para\'{\i}ba, Brazil}
\author{B. A. Malomed}
\affiliation{Department of Physical Electronics, School of Electrical Engineering,
Faculty of Engineering, Tel Aviv University, Tel Aviv 69978, Israel}

\begin{abstract}
Extending the recent work on models with spatially nonuniform
nonlinearities, we study bright solitons generated by the nonpolynomial
self-defocusing (SDF) nonlinearity in the framework of the one-dimensional
(1D) Mu\~{n}oz-Mateo - Delgado (MM-D) equation (the 1D reduction of the
Gross-Pitaevskii equation with the SDF nonlinearity), with the local
strength of the nonlinearity growing at $|x|\rightarrow \infty $ faster than
$|x|$. We produce numerical solutions and analytical ones, obtained by means
of the Thomas-Fermi approximation (TFA), for nodeless ground states, and for
excited modes with $\ 1$, $2$, $3,$ and $4$ nodes, in two versions of the
model, with steep (exponential) and mild (algebraic) nonlinear-modulation
profiles. In both cases, the ground states and the single-node ones are
completely stable, while the stability of the higher-order modes depends on
their norm (in the case of the algebraic modulation, they are fully
unstable). Unstable states spontaneously evolve into their stable
lower-order counterparts.
\end{abstract}

\maketitle

\emph{Introduction} - The experimental realization of the Bose-Einstein
condensates (BECs) of dilute atomic gases \cite%
{AndersonSCi95,BradleyPRL95,DavisPRL95} allows the investigation of a great
many of fascinating phenomena, such as the Anderson localization of matter
waves \cite{BillyNAT08,RoatiNAT08}, production of bright \cite%
{KhaykovichSCi02,StreckerNAT02,Rb85,Cornish} and dark solitons \cite%
{BurgerPRL99}, dark-bright complexes \cite{dark-bright}, vortices \cite%
{MatthewsPRL99} and vortex-antivortex dipoles \cite%
{Anderson,Hall,Bagnato,dipole}, persistent flows in the toroidal geometry
\cite{flow,flow2}, skyrmions \cite{skyrmion}, emulation of gauge fields \cite%
{gauge} and spin-orbit coupling \cite{SO}, quantum Newton's cradles \cite%
{cradle}, \textit{etc}. This subject has been greatly upheld by the use of
the Feshbach-resonance (FR) technique, i.e., the control of the strength of
the inter-atomic interactions by externally applied fields \cite%
{InouyeNAT98,CourteillePRL98,RobertsPRL98}, which opens the possibility to
implement sophisticated nonlinear patterns. In particular, the management of
localized solutions of the Gross-Pitaevskii equation (GPE) \cite{GPE} by
means of the spatially inhomogeneous nonlinearity, which may be created by
external nonuniform fields that induce the corresponding FR landscape, has
attracted a great deal of interest in theoretical studies \cite%
{Malomed06,KartashovRMP11,BBPRL08,SalasnichPRA09,AvelarPRE09,CardosoNA10,AvelarPRE10,CardosoPLA10,CardosoPLA10-2,CardosoPRE12}%
.

In this vein, the existence of bright solitons in systems with \emph{purely
repulsive}, alias self-defocusing (SDF) nonlinearity, in the absence of
external linear potentials, was recently predicted \cite{BorovkovaPRE11}.
This result is intriguing because the existence of such solutions, supported
by SDF-only nonlinearities, without the help of a linear potential, was
commonly considered impossible. In the setting introduced in Ref. \cite%
{BorovkovaPRE11}, the system is described by a nonlinear Schr\"{o}dinger
(NLS) equation with the SDF cubic term, whose strength increases in space
rapidly enough towards the periphery. The discovery of bright solitons in
this setting has ushered studies of solitary modes in other models with
spatially growing repulsive nonlinearities, both local \cite%
{BorovkovaOL11,KartashovOL11,Zeng12,LobanovOL12,KartashovOL12,Young-SPRA13}
and nonlocal \cite{He}. More specifically, in Ref. \cite{BorovkovaOL11} it
was demonstrated that spatially inhomogeneous defocusing nonlinear
landscapes modulated as $1+|r|^{\alpha }$, with $\alpha >D$ in the space of
dimension $D$, support stable fundamental and higher-order bright solitons,
as well as localized vortices, with algebraically decaying tails. Further,
it was shown in Ref. \cite{KartashovOL11} that bimodal systems with a
similar spatial modulation of the SDF cubic nonlinearity can support stable
two-component solitons, with overlapping or separated components. Work \cite%
{Zeng12} addressed the possibility of supporting stable bright solitons in
1D and 2D media by the SDF quintic term with a spatially growing
coefficient. In work \cite{LobanovOL12}, it was predicted that a practically
relevant setting, in the form of a photonic- crystal fiber whose strands are
filled by an SDF nonlinear medium, gives rise to stable bright solitons and
vortices. Asymmetric solitons and domain-wall patterns, supported by
inhomogeneous defocusing nonlinearity were reported in Ref. \cite%
{KartashovOL12}. Going beyond the limits of BEC and optics, in Ref. \cite%
{Young-SPRA13} self-trapped ground states were predicted to occur in a
spin-balanced gas of fermions with repulsion between the spinor components,
provided that the repulsion strength grows from the center to periphery, in
the combination with the usual harmonic-oscillator trapping potential acting
in one or two transverse directions. A very recent result demonstrates that
the 2D isotropic or anisotropic nonlinear potential, induced by the strength
of the local self-repulsion growing $\sim r^{4}$, can efficiently trap
fundamental solitons and vortices with topological charges $1$ and $2$ in
the dipolar BEC, with the long-range repulsion between dipoles polarized
perpendicular to the system's confinement plane \cite{Kishor}.

In the present work, we address the existence of stable bright solitons in
the framework of the nonpolynomial Mu\~{n}oz-Mateo-Delgado (MM-D) equation
\cite{MateoPRA08,Delgado2} (see also Ref. \cite{Gerbier}), which is a
one-dimensional (1D) reduction of the full three-dimensional GPE for
cigar-shaped condensates with repulsive interatomic interactions. Because of
the repulsive sign of the intrinsic nonlinearity, the MM-D equation is
drastically different from the nonpolynomial NLS equation derived as a
result of the dimensional reduction for the self-attractive BEC \cite{Luca}.
In this work, we consider the spatially modulated nonpolynomial nonlinearity
whose strength increases rapidly enough towards the periphery, similar to
what was originally introduced in Ref. \cite{BorovkovaPRE11} for the cubic
SDF nonlinearity. Results are obtained analytically by means of the
Thomas-Fermi approximation (TFA), and in a numerical form.

\emph{The theoretical model} - We start with the GPE written in 3D as
\begin{equation}
i\hbar \frac{\partial \Psi }{\partial t}=-\frac{\hbar ^{2}}{2m}\nabla
^{2}\Psi +g(\mathbf{r})N|\Psi |^{2}\Psi +\frac{1}{2}m\omega _{\perp
}^{2}\left( y^{2}+z^{2}\right) \Psi ,  \label{GP3D}
\end{equation}%
where $\Psi (\mathbf{r},t)$ is the mean-field wave function, $\nabla ^{2}$
is the Laplacian, $g(\mathbf{r})$ is the spatially-dependent local
coefficient of the self-repulsive nonlinearity, $\omega _{\perp }$ is the
strength of the harmonic-oscillator (HO) trapping potential applied in the
transverse plane, $\left( y,z\right) $, while $m$ and $N$ are the atomic
mass and the number of particles, respectively. In Refs. \cite{MateoPRA07}
and \cite{MateoPRA08} it has been shown that the effective 1D equation
governing the axial dynamics of cigar-shaped condensates with the repulsive
interatomic interactions can be derived as a reduction of the 3D equation (%
\ref{GP3D}):
\begin{equation}
i\hbar \frac{\partial \psi }{\partial t}=-\frac{\hbar ^{2}}{2m}\frac{%
\partial ^{2}\psi }{\partial x^{2}}+\hbar \omega _{\bot }\sqrt{1+4a(x)N|\psi
|^{2}}\psi ,  \label{GP1D}
\end{equation}%
with $a(x)>0$ being the $s$-wave scattering length, whose dependence on
axial coordinate $x$ may be imposed by means of the FR management. The
corresponding 3D wave function is approximated by the factorized ansatz,$\
\Psi =\psi (x,t)\Phi (\mathrm{r}_{\bot },n_{1}(x,t))$, where $n_{1}$ is the
axial density, $n_{1}=N\int \int \mathrm{d}y\mathrm{d}z\mathrm{~}|\Psi
\left( x,y,z\right) |^{2}$, and \textbf{$\Phi (\mathrm{r}_{\bot },$}$n_{1})$
is the transverse wave function satisfying equation [see Eq. (15) of Ref.
\cite{MateoPRA08})]
\begin{equation}
\left( -\frac{1}{2}\overline{\nabla }^{2}+\frac{1}{2}\overline{r}_{\bot
}^{2}+4\pi an_{1}|\overline{\Phi }|^{2}\right) \overline{\Phi }=\overline{%
\mu }_{\bot }\overline{\Phi },  \label{MD_axial}
\end{equation}%
written in scaled variables $\overline{r}_{\bot }=r_{\bot }/a_{\bot }$, $%
\overline{\Phi }=a_{\bot }\Phi $, and $\overline{\mu }_{\bot }=\mu _{\bot
}/\hbar \omega _{\bot }$. Here $\mu _{\bot }=\mu _{\bot }(n_{1})$ is the
local chemical potential, and $a_{\bot }=\sqrt{\hbar /m\omega _{\bot }}$ is
the confinement length in the transverse direction. Equation (\ref{MD_axial}%
) admits explicit approximate solutions in the limit cases of $an_{1}\ll 1$
and $an_{1}\gg 1$, treating $\overline{\Phi }$, severally, as the Gaussian
ground state of the HO potential, or the TFA wave function \cite{MateoPRA08}%
.

Finally, Eq. (\ref{GP1D}) is transformed into a scaled form,
\begin{equation}
i\frac{\partial \varphi }{\partial t}=-\frac{1}{2}\frac{\partial ^{2}\varphi
}{\partial x^{2}}+\sqrt{1+\sigma (x)|\varphi |^{2}}\varphi ,  \label{1D}
\end{equation}%
where $\sigma (x)\equiv 4a(x)/a_{\bot }$. Stationary solutions to Eq. (\ref%
{1D}) with (longitudinal) chemical potential $\mu $ are looked for as usual,
\begin{equation}
\varphi (x,t)=\phi (x)e^{-i\mu t}.  \label{stationary}
\end{equation}

The application of the TFA, which neglects the kinetic-energy term \cite{GPE}%
, to Eq. (\ref{1D}) immediately yields
\begin{equation}
\phi _{\mathrm{TFA}}^{2}=\frac{\mu ^{2}-1}{\sigma (x)},  \label{TF1D}
\end{equation}%
provided that the chemical potential takes values $\mu >1$. In the same
approximation, the norm (scaled number of atoms) of the condensate is%
\begin{equation}
N_{\mathrm{TFA}}=\left( \mu ^{2}-1\right) \int_{-\infty }^{+\infty }\frac{dx%
}{\sigma (x)},  \label{N}
\end{equation}%
provided that $\sigma (x)$ grows at $|x|\rightarrow \infty $ faster than $|x|
$, to secure the convergence of the integral. In fact, the latter condition
is the exact one which is necessary and sufficient for the existence of
physically relevant self-trapped modes in the MM-D equation.

The results produced by the TFA in the form of Eq. (\ref{TF1D}) were used as
the initial guess for finding numerically exact stationary solutions by
means of the well-known method of the imaginary-time integration, in the
framework of which the convergence of stationary solutions may be related to
their stability against small perturbations in real time \cite{Yang10}. As
characteristic examples, we use two different axial nonlinearity-modulation
profiles,
\begin{equation}
\sigma _{\mathrm{A}}(x)=\cosh ^{2}(2x);~~\sigma _{\mathrm{B}}(x)=1+x^{6},
\label{AB}
\end{equation}%
cf. Refs. \cite{BorovkovaPRE11}-\cite{Zeng12}. Below, these two profiles are
referred to as Cases A and B, respectively.

The stability of the self-trapped modes was investigated in the framework of
the linearized equations, taking the perturbed solutions as%
\begin{equation}
\varphi =\left\{ \phi (x)+\left[ v(x)+w(x)\right] e^{\lambda t}+\left[
v^{\ast }(x)-w^{\ast }(x)\right] e^{-\lambda ^{\ast }t}\right\} e^{i\mu t},
\label{LS}
\end{equation}%
where $v(x)$ and $w(x)$ are small perturbations, and $\lambda $ is the
respective eigenvalue. The ensuing linear-stability eigenvalue problem was
solved by means of the Fourier Collocation Method, as described in Ref. \cite%
{Yang10}.

First, we have analyzed the stability for the ground-state solutions
produced by the TFA (used as the initial guess for the numerical stationary
solutions). It has been found that they are stable for all $\mu >1$, see
further details below. It is also worthy to note that, as it follows from
Eq. (\ref{N}) and corroborated below by numerical results, the families of
the ground-state modes obey the anti-Vakhitov-Kolokolov (anti-VK) criterion,
$dN/d\mu >0$, which is a necessary stability condition for localized modes
supported by SDF nonlinearities \cite{anti} (the VK criterion per se,
relevant to the usual solitons supported by the self focusing nonlinearity,
\cite{VakhitovRQE73,BergePR98}, has the opposite form, $dN/d\mu <0$).

\emph{Numerical Results} - The integration in imaginary time was carried out
by means of the split-step code, which was composed so as to restore the
original norm of the solution at the end of each step of marching forward in
imaginary time. The dispersive part of Eq. (\ref{1D}) was handled by means
of the Crank-Nicolson algorithm with spatial and temporal steps $\Delta
x=0.04$ and $\Delta t=0.001$.

To find higher-order stationary solutions with nodes, the Gram-Schmidt
orthogonalization was performed at the end of each time step. The
ground-state and higher-order solutions were thus obtained, using the
Hermite-Gaussian input profiles of orders $n=0,1,2,3,4.$ To check the
correctness of the stationary solutions, we have also reproduced them by
means of the standard relaxation method, concluding that the solutions
obtained by dint of both techniques were indistinguishable. Finally, the
stability was checked via the real-time simulations of the perturbed
evolution of input profiles to which random noise was added at the $5\%$
amplitude level, as well as through the computation of the stability
eigenvalues for perturbed solution taken as in Eq. (\ref{LS}).

\begin{figure}[tb]
\centering
\par
\includegraphics[width=0.6\columnwidth]{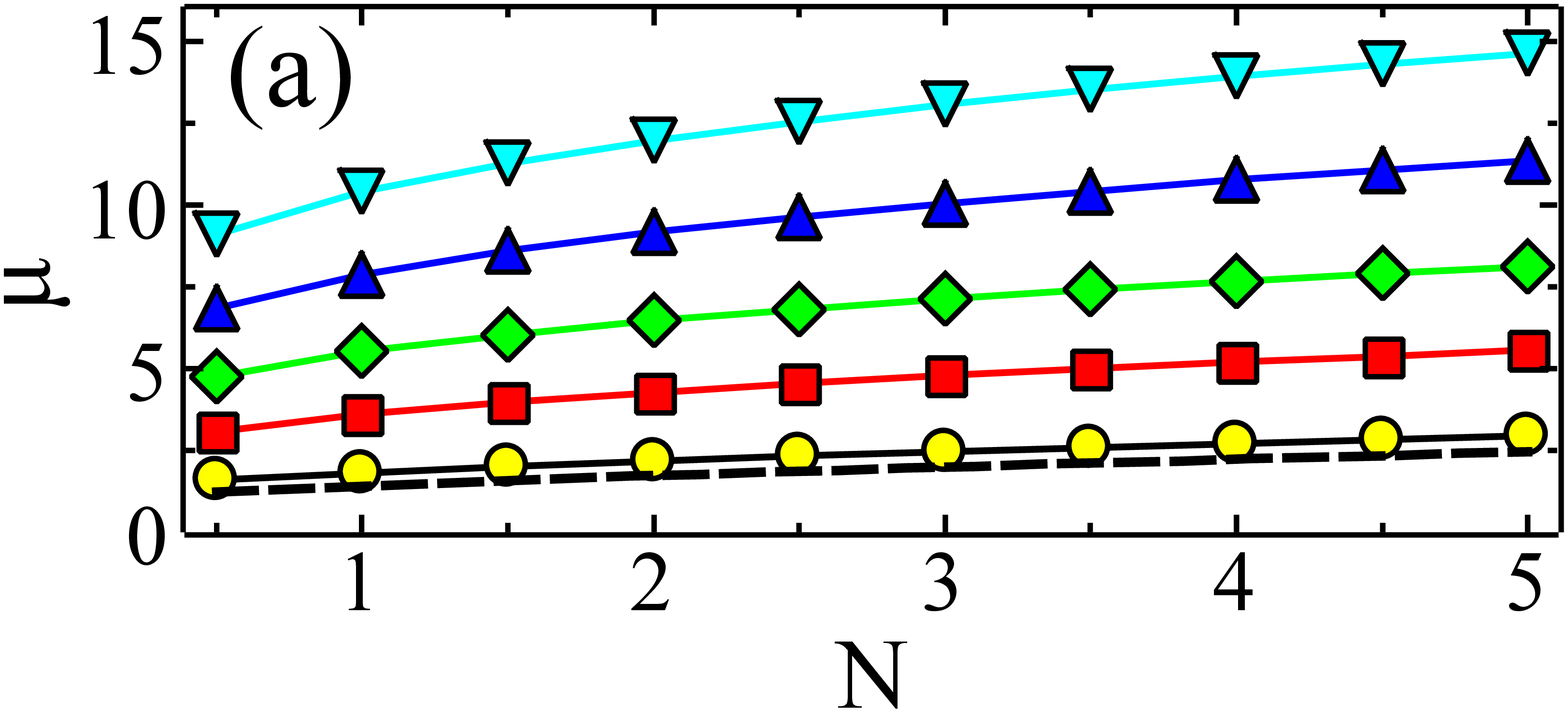} 
\includegraphics[width=0.38\columnwidth]{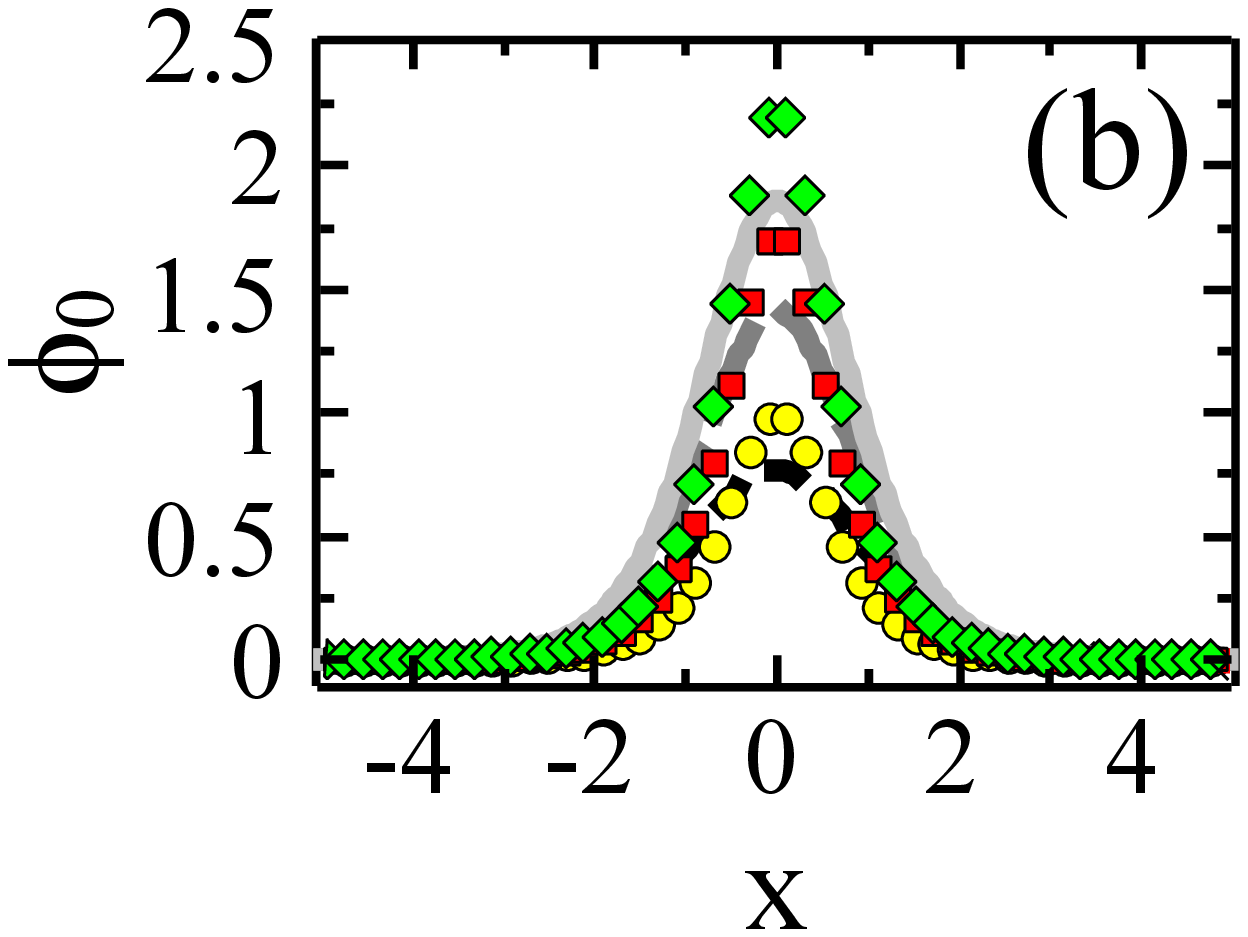} 
\includegraphics[width=0.38\columnwidth]{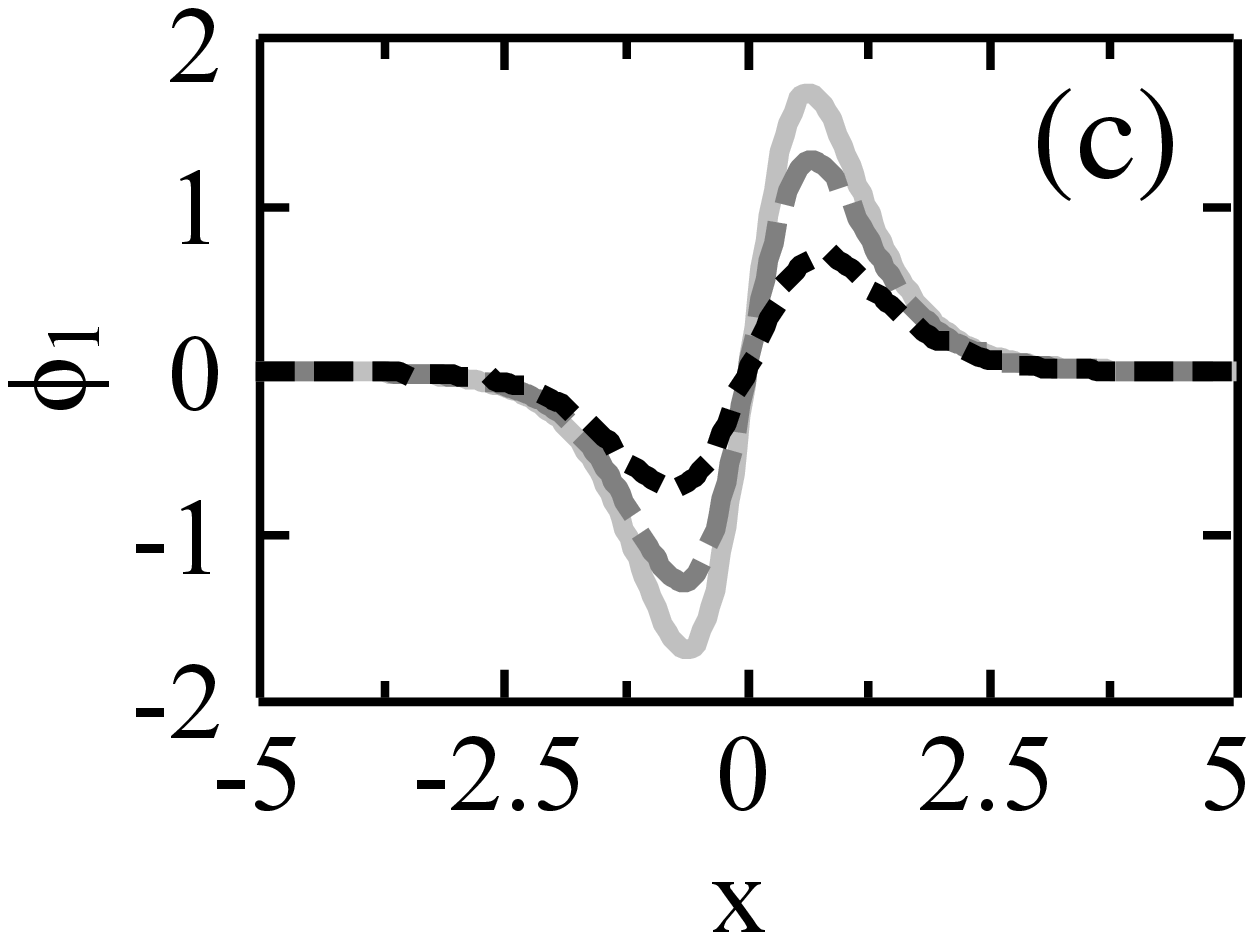}\hfil
\includegraphics[width=0.38\columnwidth]{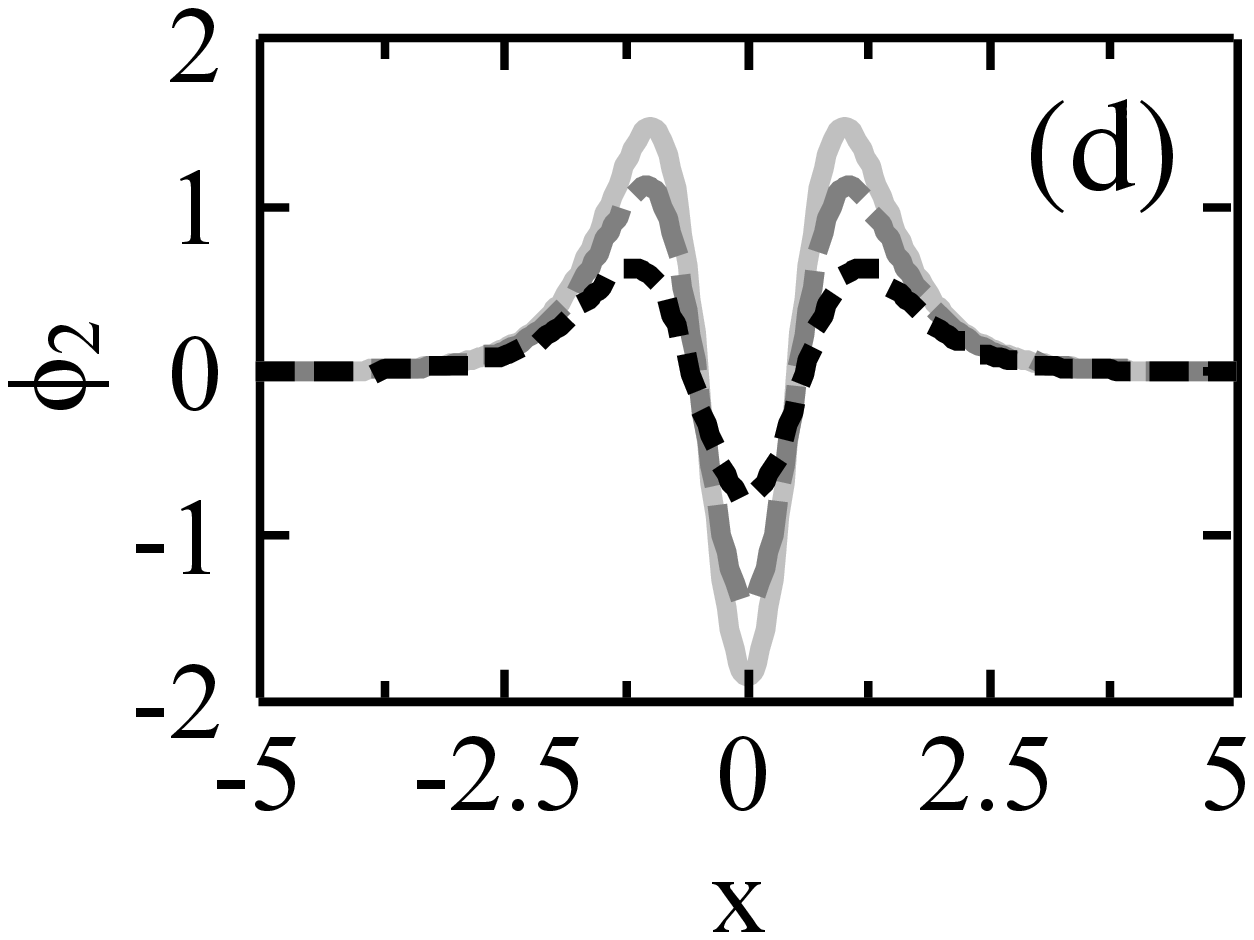} 
\includegraphics[width=0.38\columnwidth]{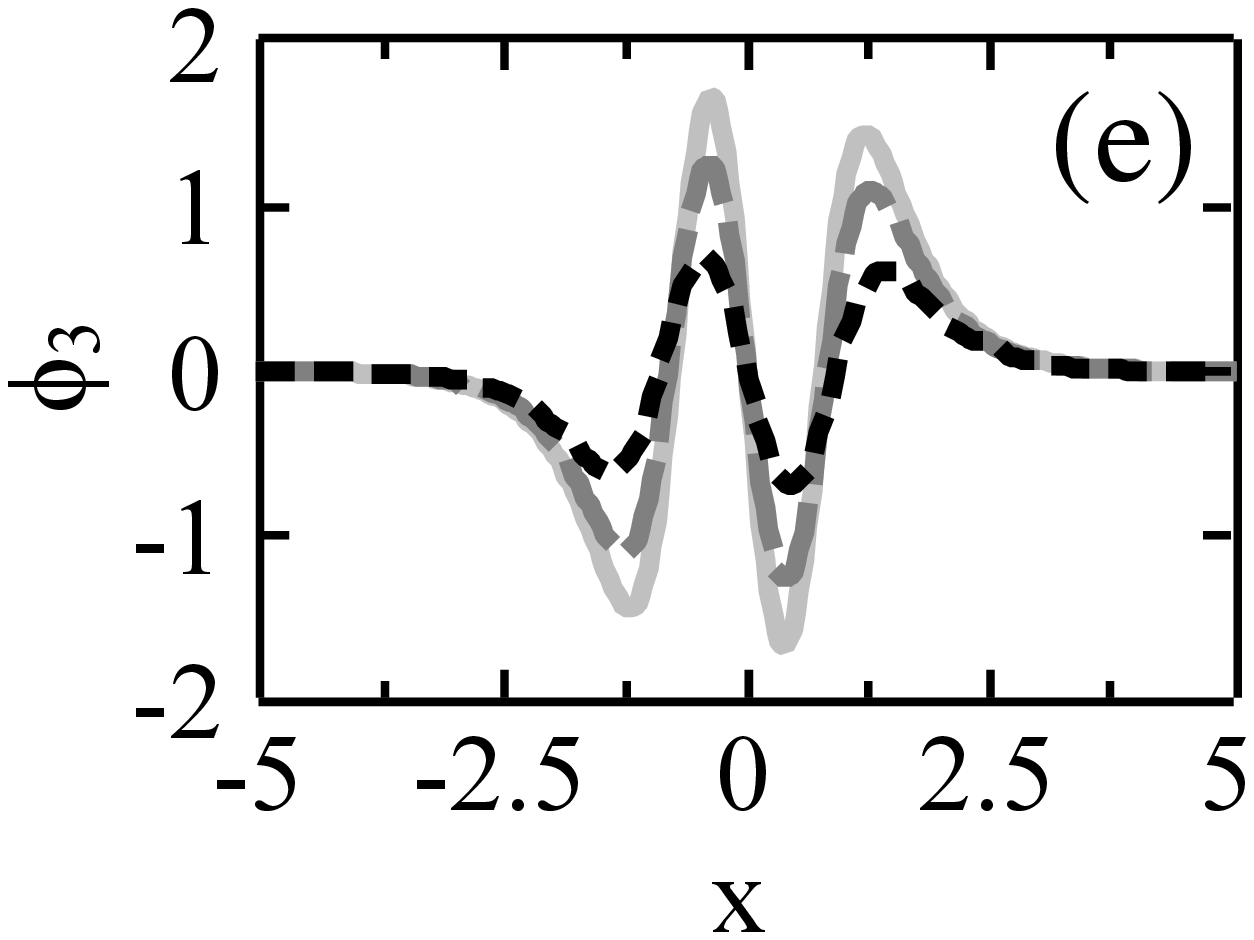}
\hfil \includegraphics[width=0.38\columnwidth]{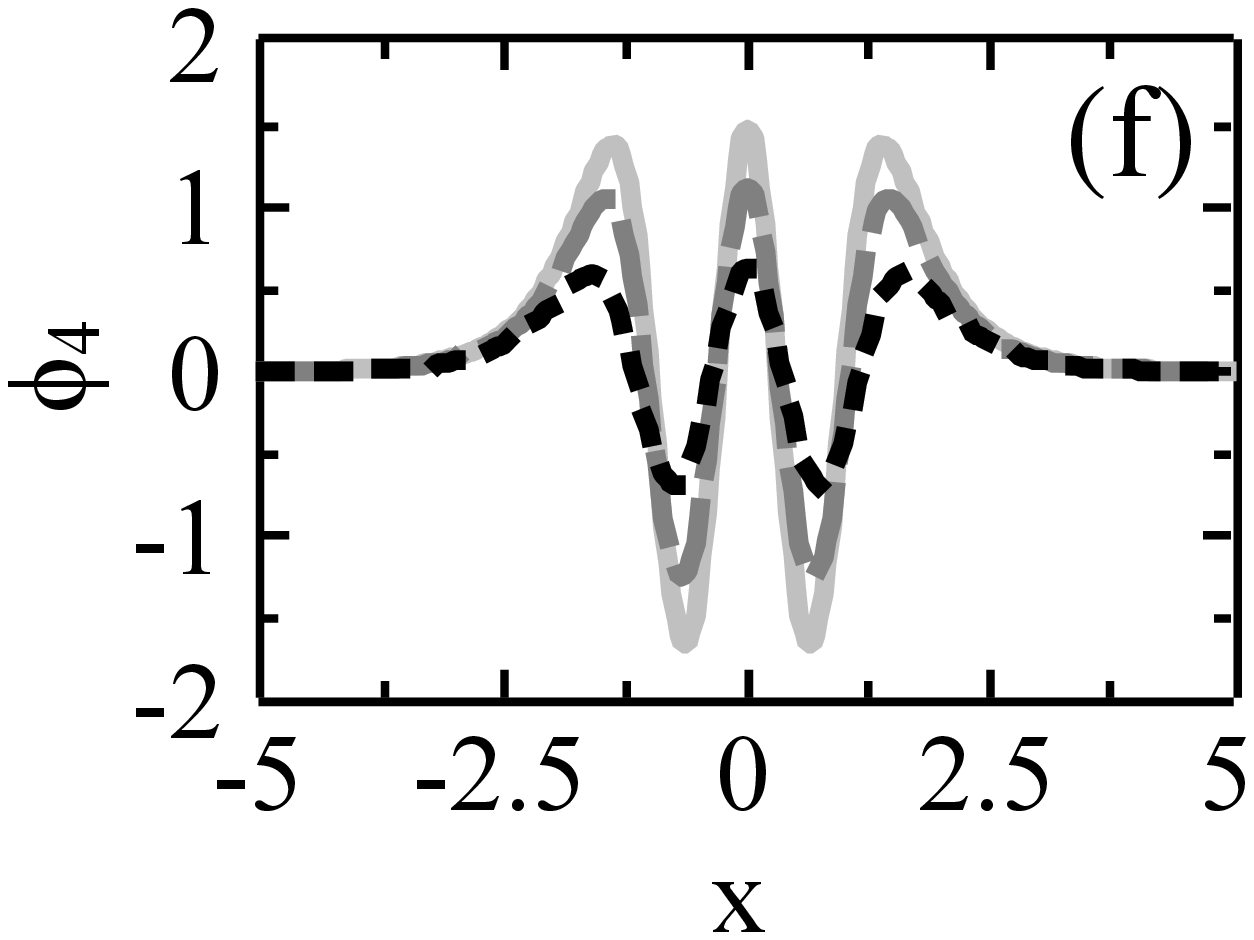}
\caption{(Color online) The model with the nonlinearity-modulation function $%
\protect\sigma _{\mathrm{A}}(x)=\cosh ^{2}\left( 2x\right) $. (a) Chemical
potential $\protect\mu $ versus norm $N$ for the ground-state solution, $%
\protect\phi _{0}$ (the yellow curve with circles), the first excited state,
$\protect\phi _{1}$ (the red curve with box-shaped symbols), the
second-order mode, $\protect\phi _{2}$ (the green curve with diamond
symbols), the third-order mode, $\protect\phi _{3}$ (the blue curve with
triangles), and the fourth-order mode, $\protect\phi _{4}$ (the cyan curve
with inverted triangles). The dashed line shows the TFA prediction for the
ground state, see Eqs. (\protect\ref{TF1D}) and (\protect\ref{N}). Profiles
of stationary solutions produced by the imaginary-time integration method: $%
\protect\phi _{0}$ (b), $\protect\phi _{1}$ (c), $\protect\phi _{2}$ (d), $%
\protect\phi _{3}$ (e), and $\protect\phi _{4}$ (f), with norms $N=1$
(dashed lines), $N=3$ (long-dashed lines), and $N=5$ (solid lines). In panel
(b), the corresponding TFA-predicted shapes are shown by chains of circles ($%
N=1$), boxes ($N=3$), and diamonds ($N=5$).}
\label{FB1}
\end{figure}

\emph{Case A}) As said above, the exponential modulation profile of the
nonlinearity coefficient, corresponding to $\sigma _{\mathrm{A}}(x)$ in Eq. (%
\ref{AB}), is suggested by its counterpart that was used with the quintic
nonlinearity in Ref. \cite{Zeng12}. In Fig. \ref{FB1}(a) we display the
relation between the chemical potential $\mu $ and norm $N$ for stationary
solutions $\phi _{k}$ of different orders (number of nodes) $k$, obtained
with this profile. Typical examples of the stationary modes are shown in
panels \ref{FB1}(b-f).

All the solution branches satisfy the above-mentioned anti-VK criterion. We
have checked that, in agreement with this fact, the ground-state modes, $%
\phi _{0}$, are stable for all $\mu >1$. However, for higher-order modes
this criterion is necessary but not sufficient for the stability. We have
thus found that the first excited state, $\phi _{1}$, is fully stable, while
higher-order ones, $\phi _{2,3,4}$ are stable only in specific regions (we
have checked this up to the value of the total norm $N=15$). This trend (the
full stability of the ground and first excited states, and partial
instability of the higher-order ones) is similar to that featured by the
model with the spatially growing local strength of the cubic SDF
nonlinearity \cite{BorovkovaPRE11}.

\begin{figure}[tb]
\centering
\includegraphics[width=1.\columnwidth]{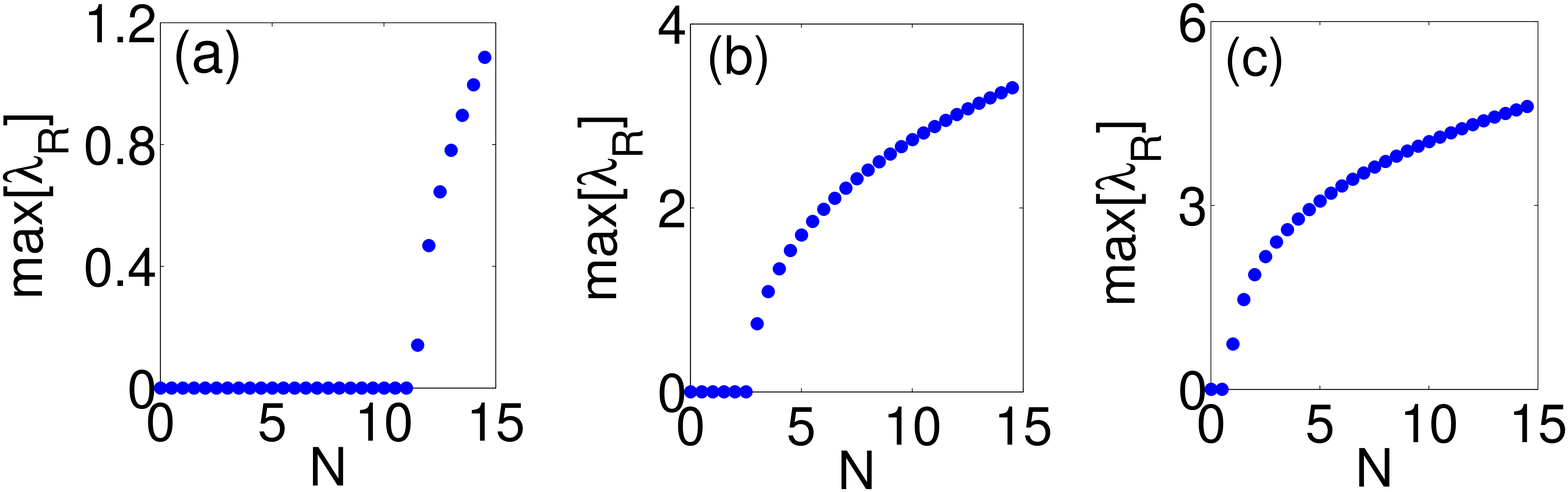}
\caption{(Color online) The model with $\protect\sigma _{\mathrm{A}%
}(x)=\cosh ^{2}\left( 2x\right) $. The largest instability growth rate, $%
\protect\lambda _{R}$, produced by the linear-stability analysis based on
Eq. (\protect\ref{LS}), versus the norm of the unperturbed solution, for the
second-order mode, $\protect\phi _{2}$ (a), third-order model, $\protect\phi %
_{3}$ (b), and the fourth-order one, $\protect\phi _{4}$ (c).}
\label{LSB}
\end{figure}

\begin{figure}[tb]
\centering \includegraphics[width=4cm]{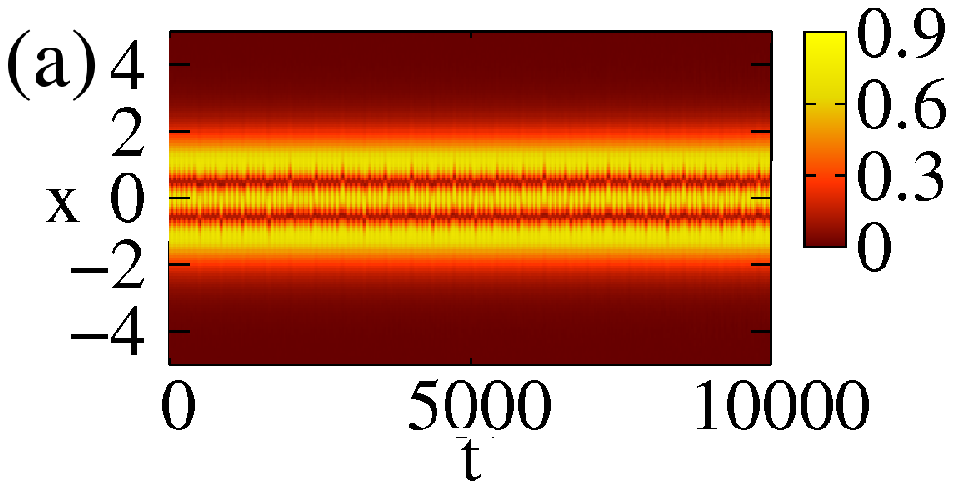} %
\includegraphics[width=4cm]{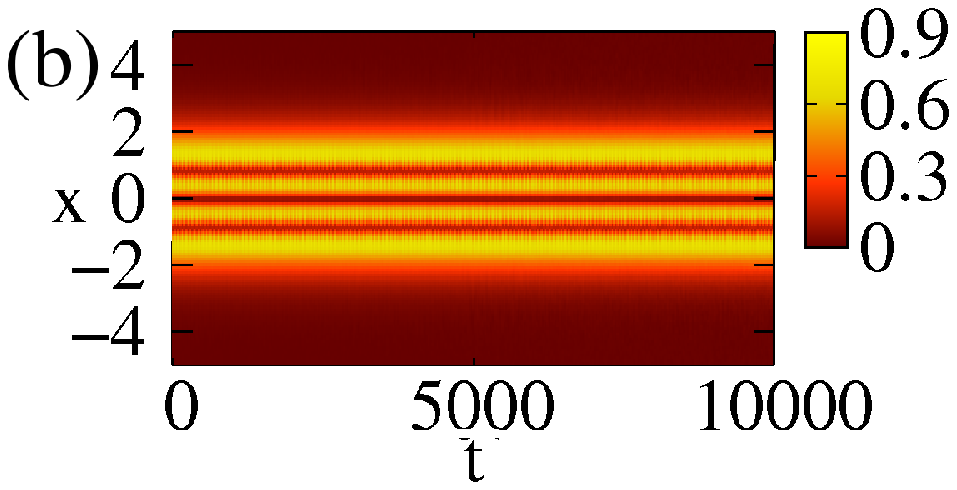} %
\includegraphics[width=4cm]{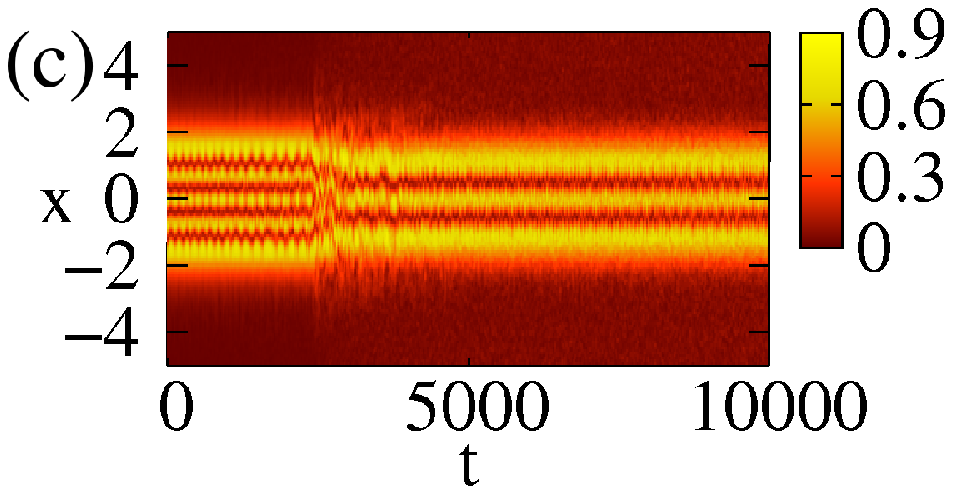}
\caption{(Color online) The model with $\protect\sigma _{\mathrm{A}%
}(x)=\cosh ^{2}\left( 2x\right) $: The real-time evolution of perturbed
modes of orders $k=2$ (a), $k=3$ (b), and $k=4$ (c) with norm $N=1$. In
panels (a) and (b) the solutions are stable, while in (c) $\protect\phi _{4}$
is unstable, decaying into $\protect\phi _{2}$.}
\label{FB2}
\end{figure}

\begin{figure}[tb]
\centering \includegraphics[width=4cm]{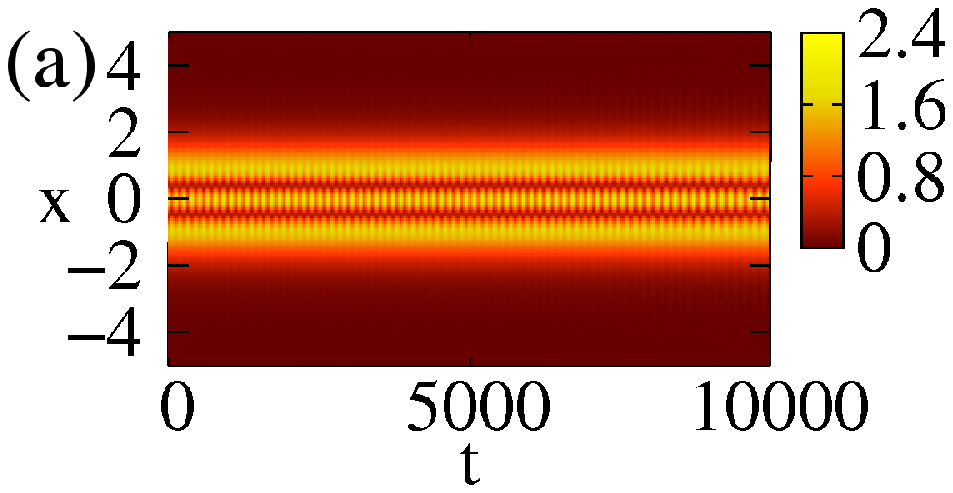} %
\includegraphics[width=4cm]{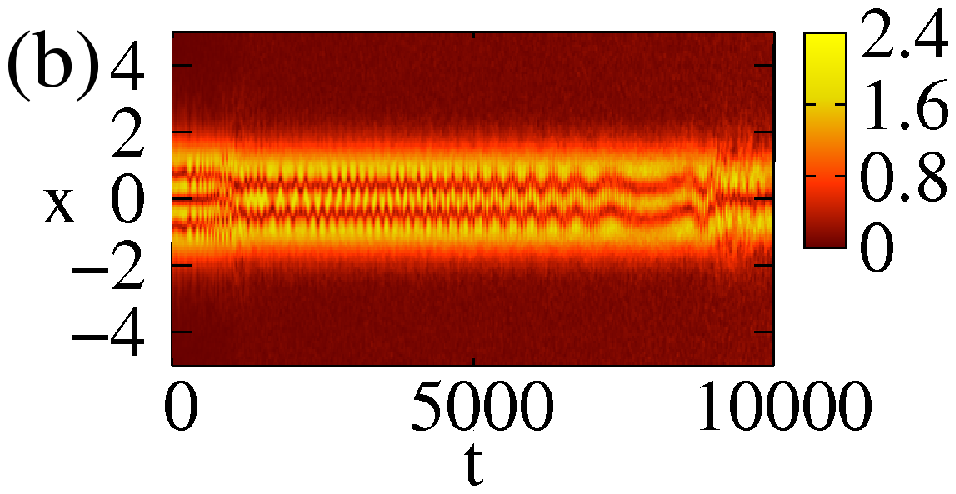} %
\includegraphics[width=4cm]{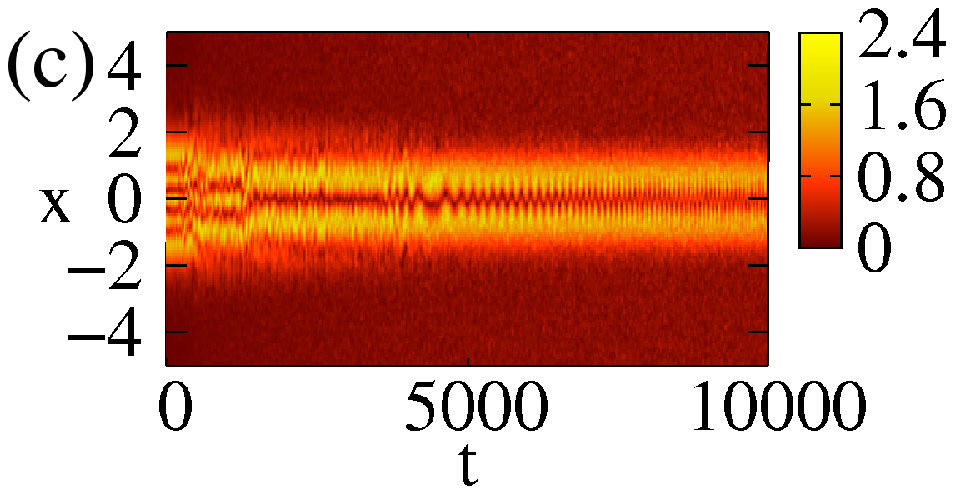}
\caption{(Color online) The same as in Fig. \protect\ref{FB2}, but for norm $%
N=5$. The solution is stable in (a), and unstable in (b) and (c).}
\label{FB3}
\end{figure}

Results of the linear-stability analysis, based on Eq. (\ref{LS}), are
presented in Fig. \ref{LSB}, which displays the largest real part of the
eigenvalue, $\lambda _{R}$, versus the norm for solutions $\phi _{2}$ (a), $%
\phi _{3}$ (b), and $\phi _{4}$ (c). It is seen that the instability sets in
with the increase of the norm. As examples, in Figs. \ref{FB2} and \ref{FB3}
we show the density profiles, $|\psi (x)|^{2}$, for $N=1$ and $N=5$,
respectively, generated in the direct simulations of the perturbed evolution
of the input states $\phi _{2}$, $\phi _{3}$, and $\phi _{4}$, with the
addition of the $5\%$ random noise. The results agree with the predictions
of the linear-stability analysis, cf. Fig. \ref{LSB}.

Direct simulations confirm the predictions of the linear-stability analysis,
as shown in Figs. \ref{FB2} and \ref{FB3}. All unstable higher-order modes
(at least, up to norm $N=15$) are spontaneously transformed into stable
modes of lower orders (with fewer nodes). In particular, the unstable $\phi
_{4}$ mode in Fig. \ref{FB3}(c) decays into $\phi _{1}$, although mode $\phi
_{2}$ is stable here too. In some other cases, the unstable mode decays
directly to the ground state, $\phi _{0}$, although the stable $\phi _{1}$
mode exists too.

%\begin{widetext}

%\end{widetext}

\begin{figure}[tb]
%\centering
\par
\includegraphics[width=0.6\columnwidth]{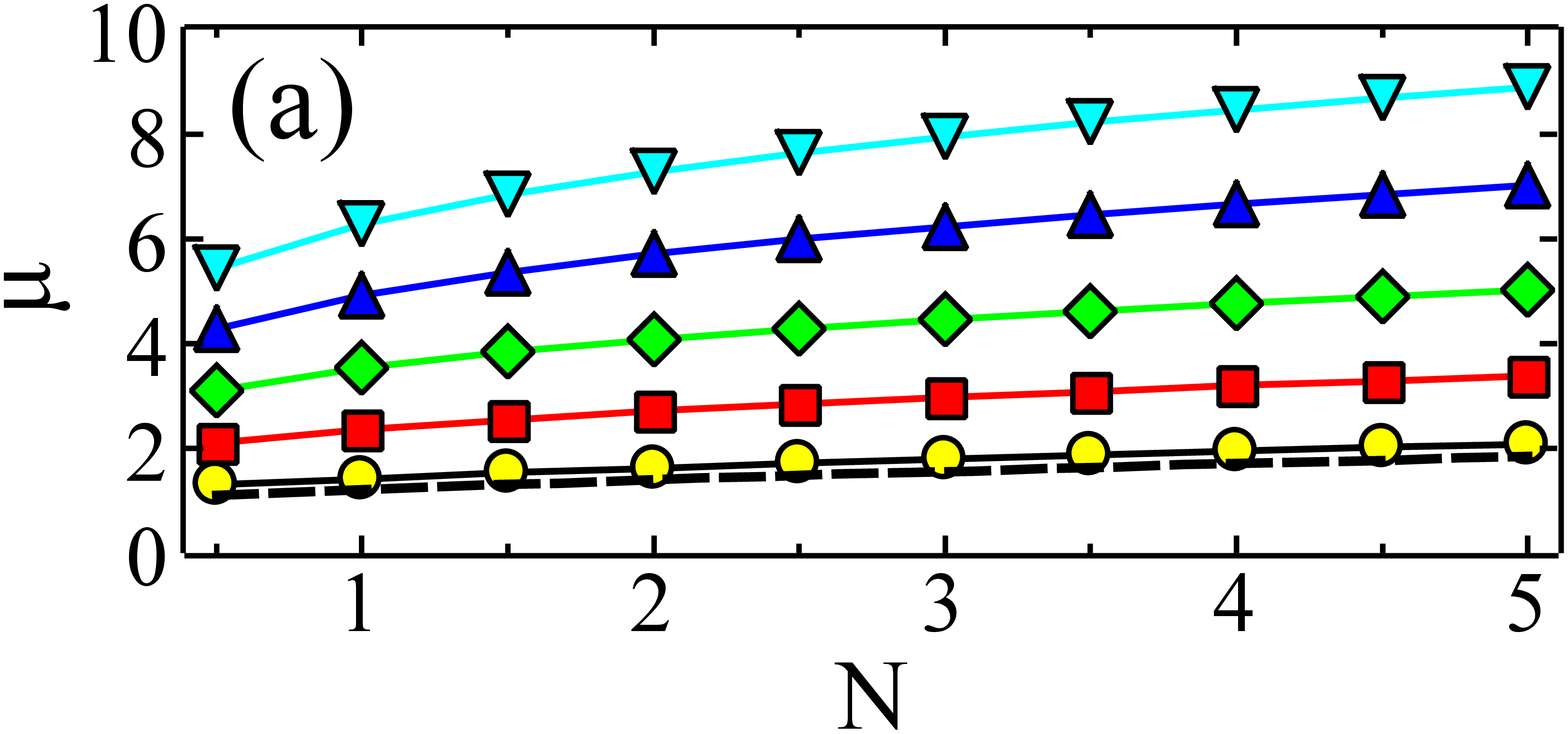} \includegraphics[width=0.38%
\columnwidth]{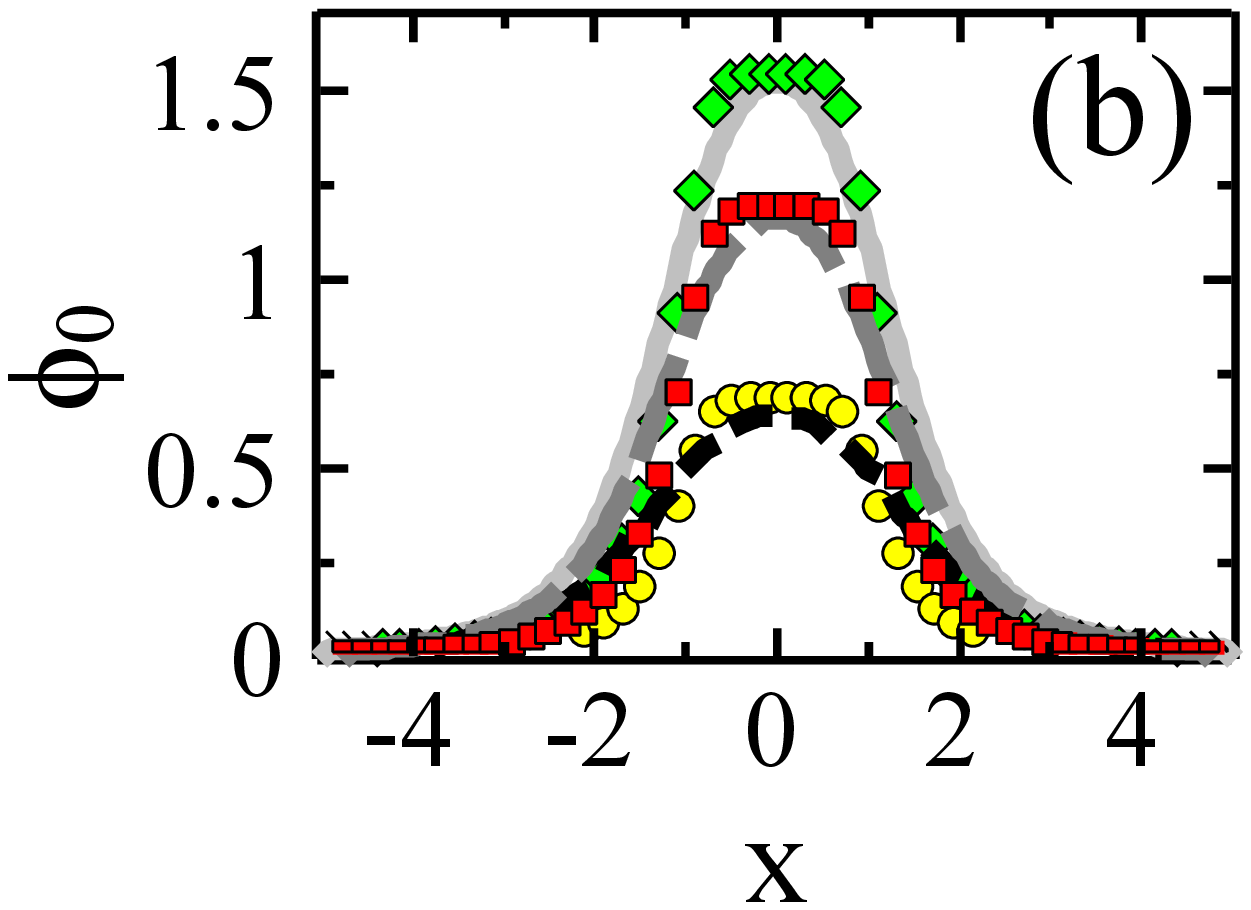} \includegraphics[width=0.38\columnwidth]{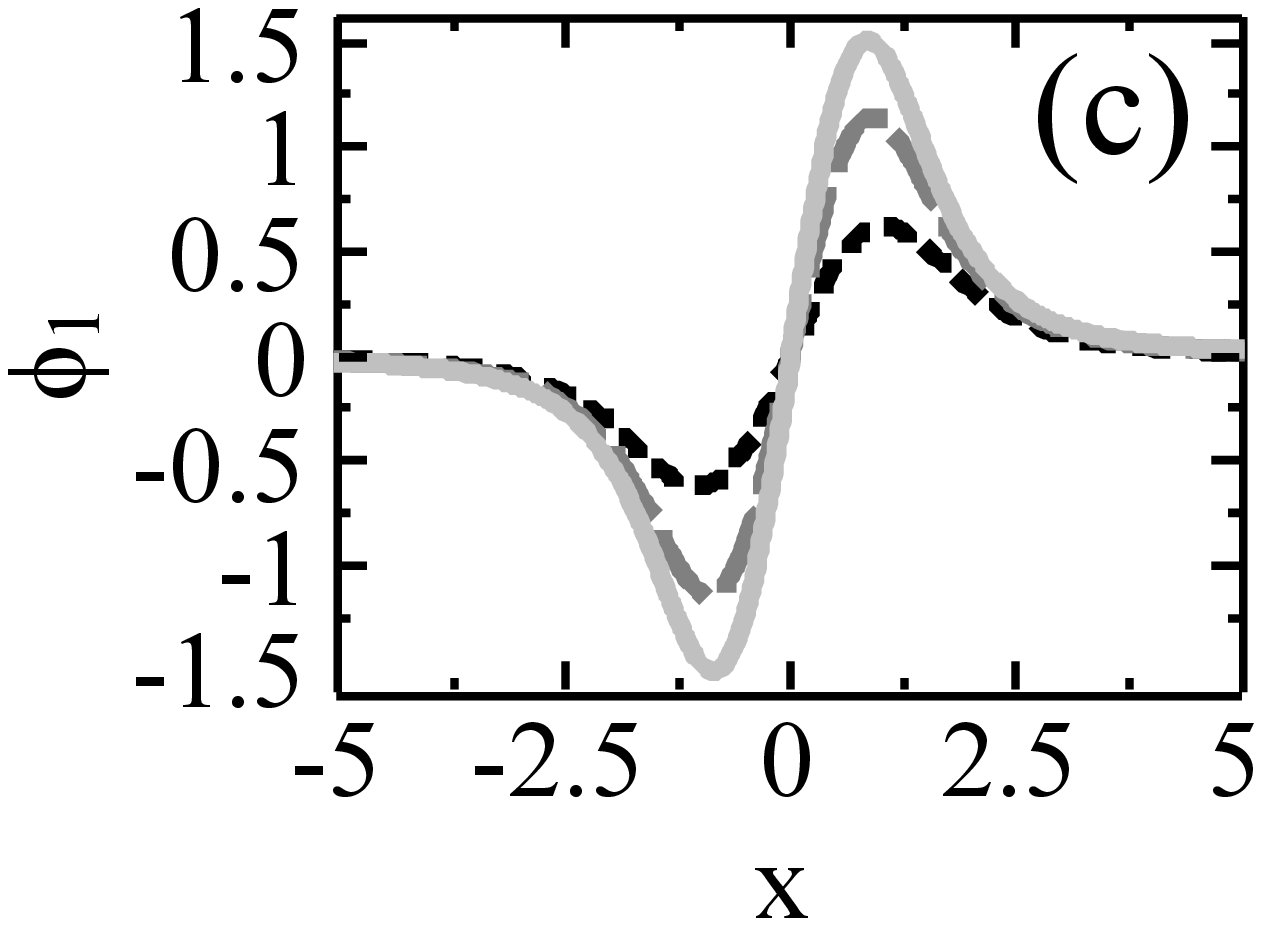}\hfil %
\includegraphics[width=0.38\columnwidth]{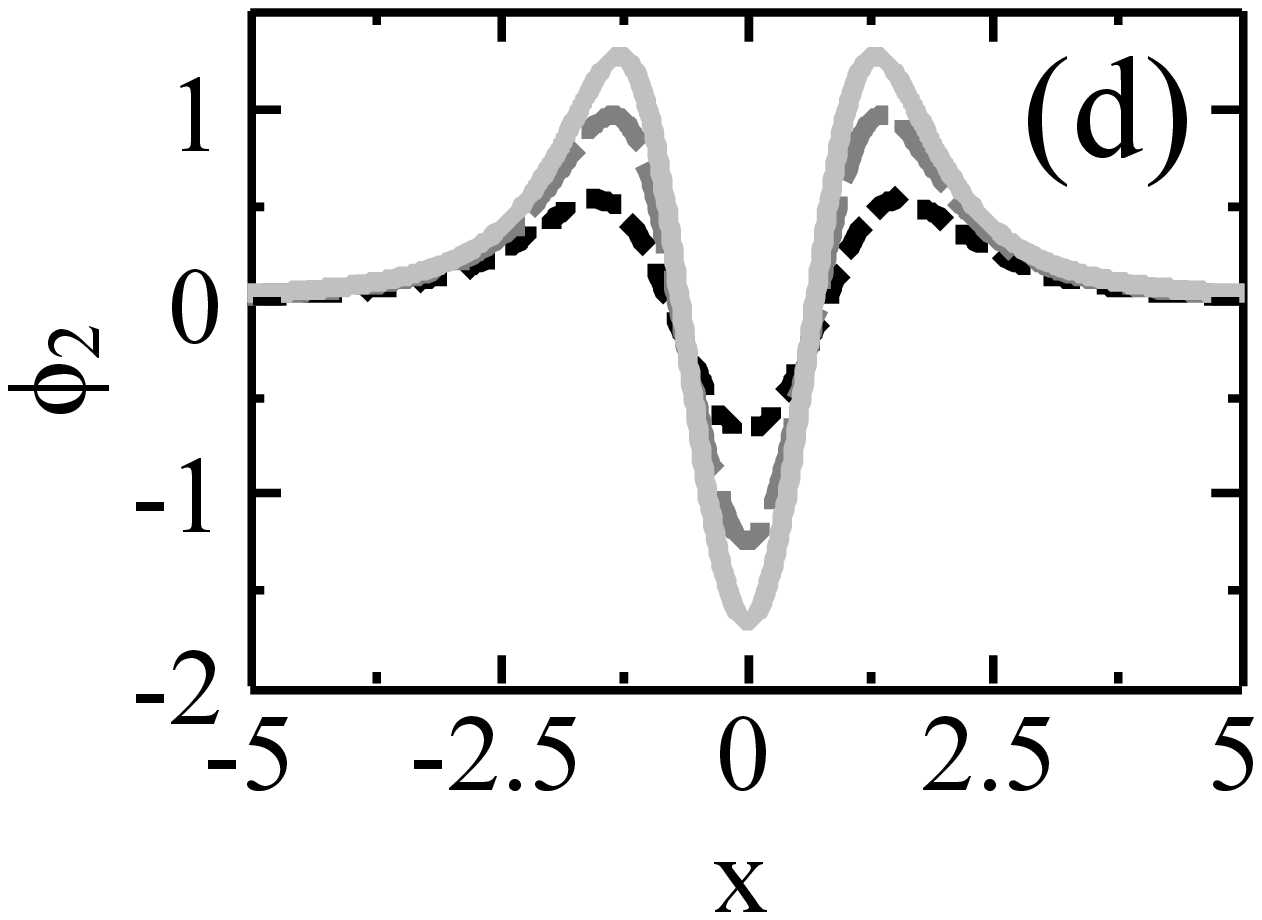} \includegraphics[width=0.38%
\columnwidth]{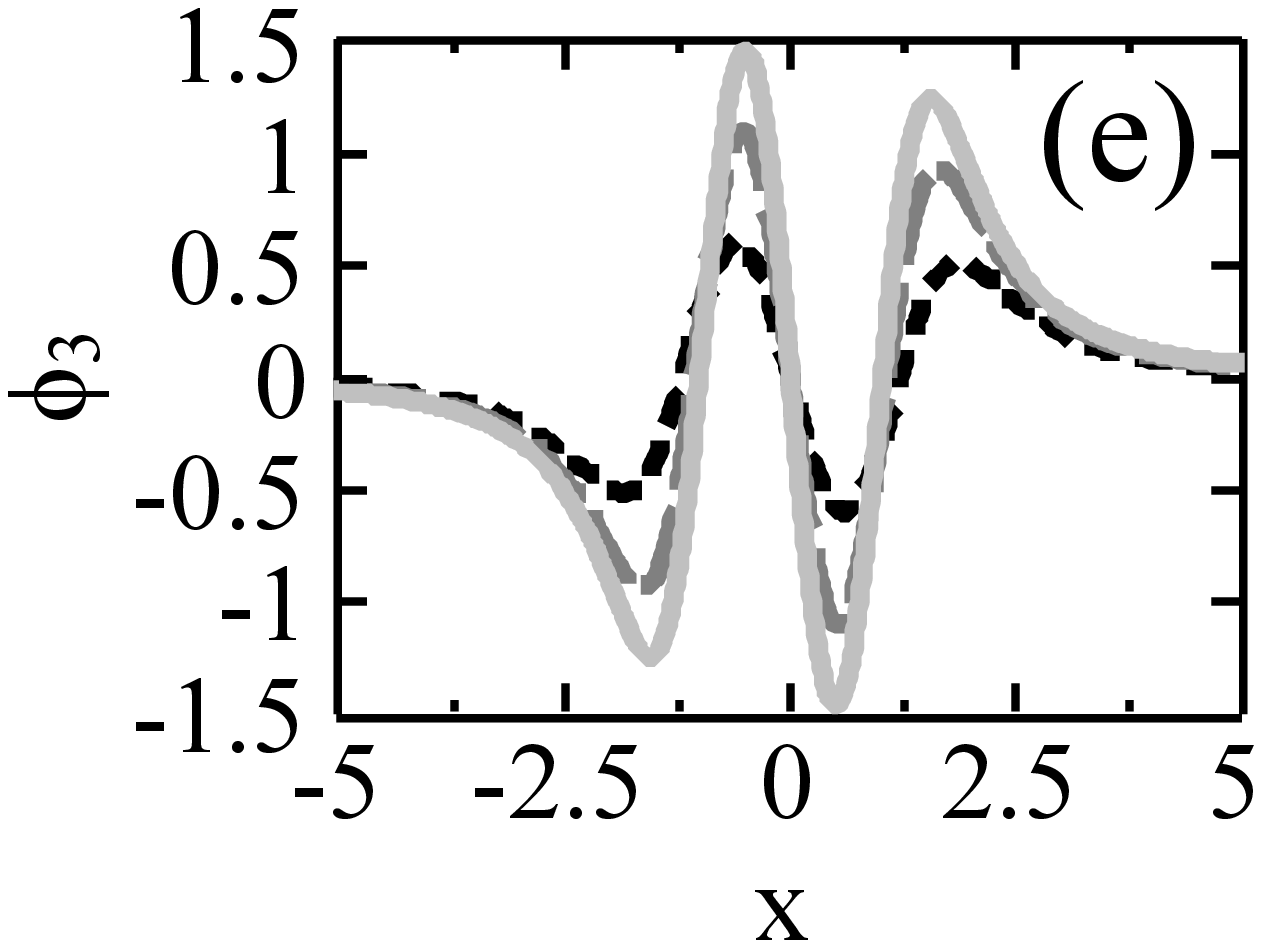}\hfil \includegraphics[width=0.38\columnwidth]{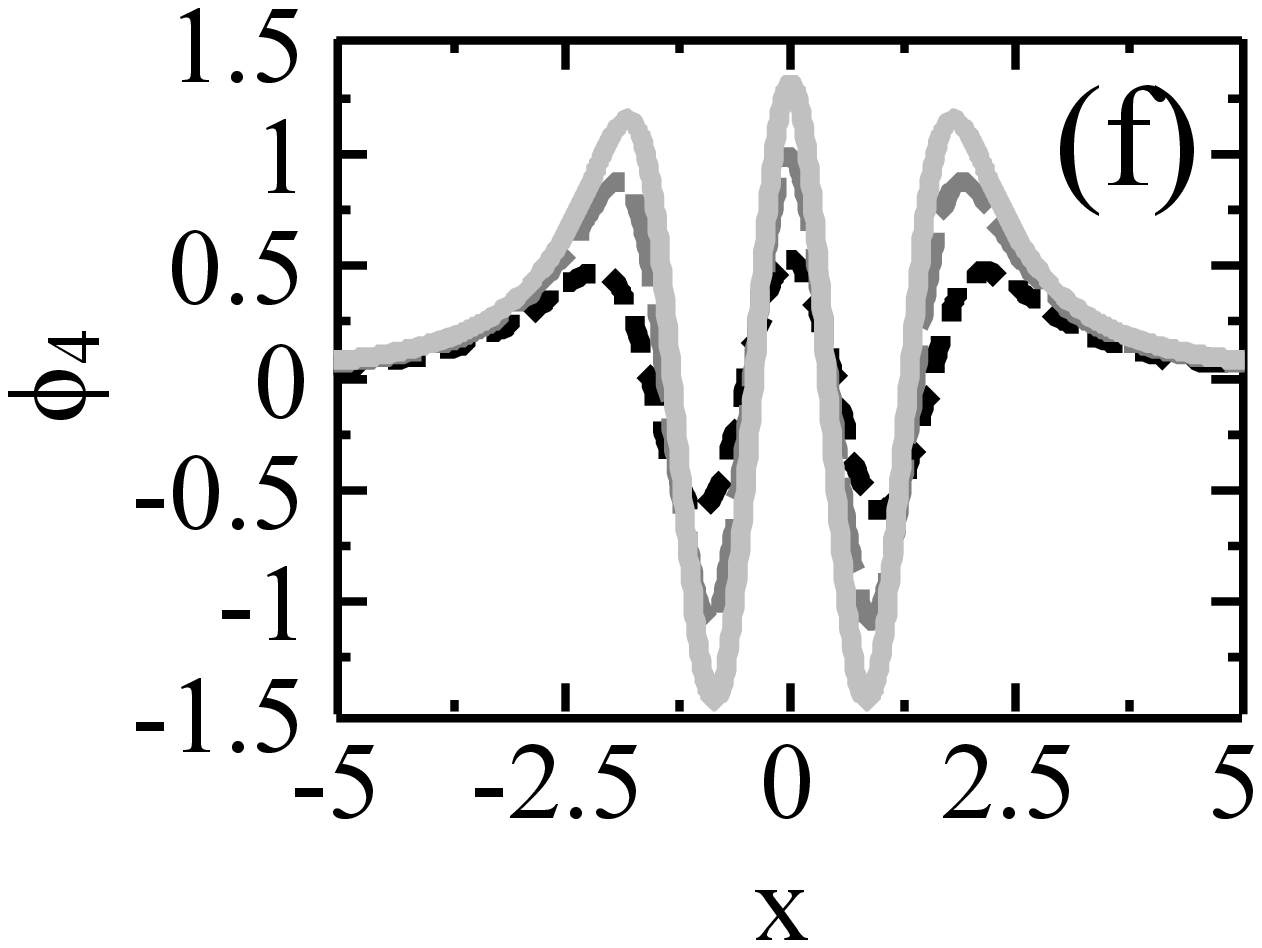}
\caption{(Color online) The same as in Fig. \protect\ref{FB1}, but for the
nonlinearity-modulation function $\protect\sigma _{\mathrm{B}}(x)=1+x^{6}$. }
\label{FA1}
\end{figure}

\emph{Case B}) The algebraic (mild) modulation of the nonlinearity
coefficient is represented by $\sigma _{\mathrm{B}}(x)$ from Eq. (\ref{AB}).
For this version of the model, chemical potential $\mu $ is shown, as a
function of norm $N$, in Fig. \ref{FA1}(a) for different modes $\phi _{k}$
[the prediction of the TFA for the ground state, given by Eq. (\ref{TF1D}),
in shown by the dashed line].

\begin{figure}[tb]
\centering \includegraphics[width=4cm]{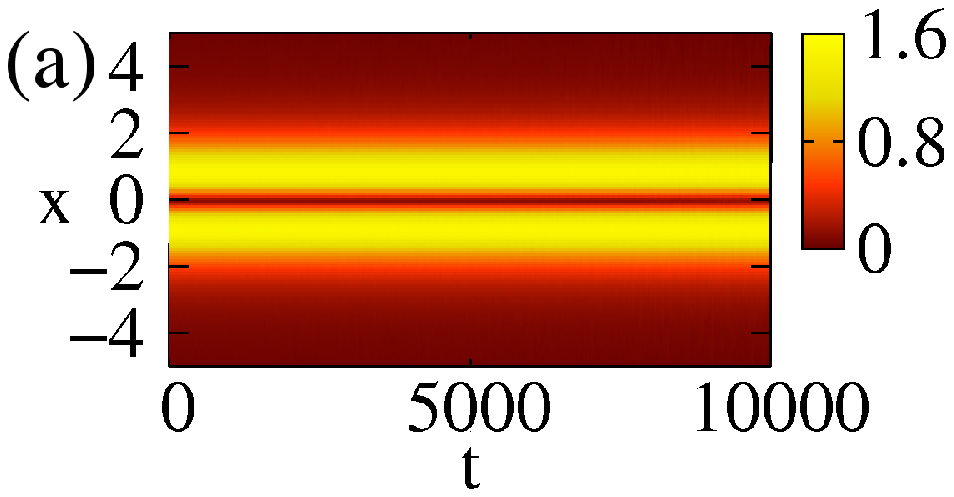} %
\includegraphics[width=4cm]{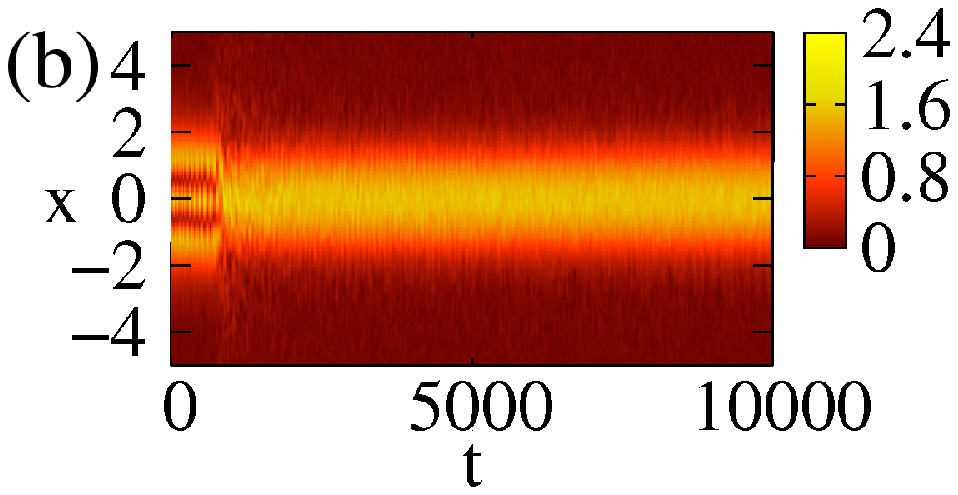} %
\includegraphics[width=4cm]{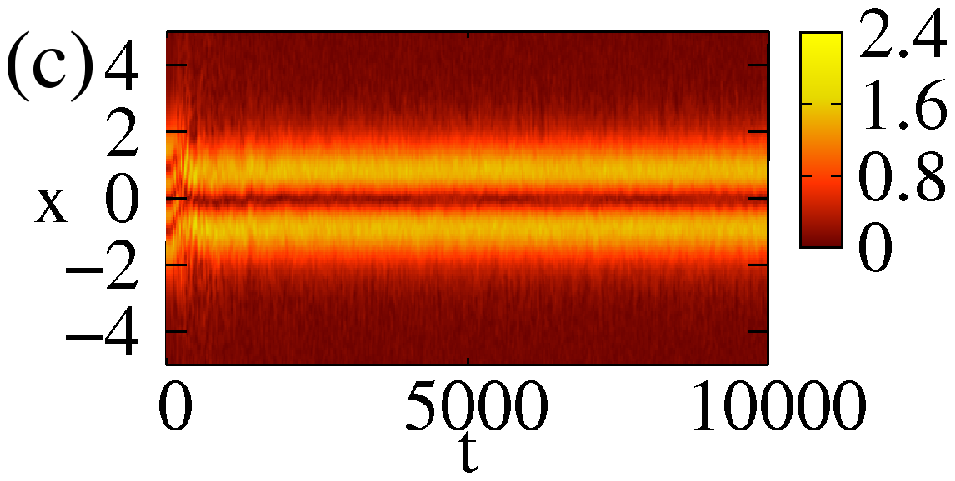} %
\includegraphics[width=4cm]{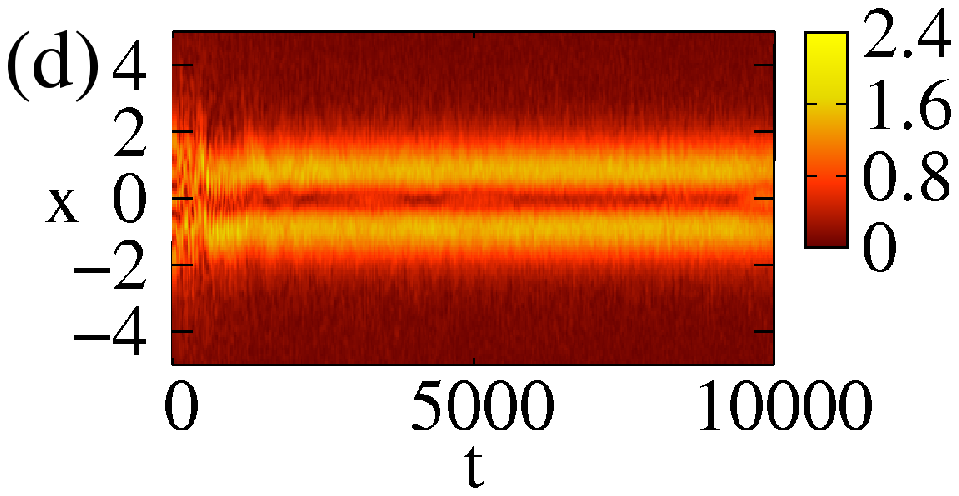}
\caption{(Color online) The same as in Fig. \protect\ref{FB3}, but for $%
\protect\sigma _{\mathrm{B}}(x)=1+x^{6}$. (a): The stable evolution of mode $%
\protect\phi _{1}$. (b), (c), and (d): Unstable evolution of modes $\protect%
\phi _{2}$, $\protect\phi _{3}$, and $\protect\phi _{4}$, respectively.}
\label{FA2}
\end{figure}

Numerical simulations confirm the stability of all the ground-state
solutions, $\phi _{0}$, for all $\mu >1$. However, the stability regions for
higher-order modes differ from the version of the model corresponding to $%
\sigma _{\mathrm{A}}(x)$ in Eq. (\ref{AB}), as concerns instability regions
for higher-order modes. In this case too, we have checked the stability for $%
N\leq 15$, concluding that $\phi _{1}$ is completely stable, while $\phi
_{2,3,4}$ are unstable at \emph{all} values of the norm. Similarly to \emph{%
Case A}, but much faster, unstable solutions decay into lower-order stable
states, $\phi _{0}$ or $\phi _{1}$. As typical examples, in Fig. \ref{FA2}
we display a stable single-node solution, $\phi _{1}$, and the spontaneous
decay of multi-node solutions into lower-order states, for $N=5$. In the
model with the modulated local strength of the cubic self-repulsive term,
the mild algebraic form of the modulation also gives rise to solitons
families which are less stable than their counterparts obtained in the model
with the steep exponential modulation, cf. Refs. \cite{BorovkovaOL11} and
\cite{BorovkovaPRE11}.\ \ \ \ \ \ \ \ \ \ \ \ \ \ \ \ \ \

\emph{Conclusion} - We have shown that the MM-D (Mu\~{n}oz-Mateo -
Delgado) equation, which is the 1D nonpolynomial reduction of the 3D
GPE with the self-repulsive cubic nonlinearity, supports stable
fundamental and higher-order self-trapped modes (``solitons"),
provided that
the local strength of the nonlinearity grows faster than $|x|$ at $%
|x|\rightarrow \infty $. We have studied in detail two
nonlinearity-modulation patterns, one steep (exponential), and one mild
(algebraic). In these models, the ground state (which was approximated
analytically by means of the TFA), and the single-node excited state are
completely stable, while the stability of higher-order (multi-node) excited
modes depends on their norm, and is different in the two models. In direct
simulations, the evolution of unstable modes always leads to their
spontaneous transformation into stable ones of lower orders.

It may be interesting to extend the analysis for fundamental solitons and
solitary vortices in the 2D setting, with the tight confinement acting in
the transverse direction, and the nonlinearity strength growing along the
radius, $r$, faster than $r^{2}$.

\emph{Acknowledgments }-\emph{\ }W.B.C, A.T.A, and D.B. thank Brazilian
agencies, CNPq, CAPES, FAPESP, and Instituto Nacional de Ci\^{e}ncia e
Tecnologia-Informa\c{c}\~{a}o Qu\^{a}ntica, for partial support. J.Z.
acknowledges support from the Natural Science Foundation of China (Project
No. 11204151). The work of B.A.M. was supported, in a part, by the
German-Israel Foundation (Grant No. I-1024-2.7/2009), and Binational
(US-Israel) Science Foundation (Grant No. 2010239).

\end{document}